\documentclass[preprint,superscriptaddress,a4paper,showpacs,amssymb]{revtex4}
\usepackage{graphicx}

\begin{document}

\title{Investigation of the charge transport through disordered organic molecular heterojunctions}
\author{H. Houili}
\affiliation{\it{Laboratoire d'Opto\'{e}lectronique des Mat\'eriaux Mol\'{e}culaires, STI-IMX-LOMM, Station 3, Ecole Polytechnique F\'{e}d\'{e}rale de Lausanne, CH-1015, Lausanne, Switzerland.}}

\author{E. Tuti\v s}
\affiliation{\it{Institute of Physics, Bijeni{\v c}ka cesta 46, P.O. Box 304, HR-10000, Zagreb, Croatia.}}

\author{I. Batisti{\'c}}
\affiliation{\it{Department of Physics, Faculty of Science, Bijeni{\v c}ka cesta 32, University of Zagreb, P.O. Box 331, HR-10002 Zagreb, Croatia.}}

\author{L. Zuppiroli}
\affiliation{\it{Laboratoire d'Opto\'{e}lectronique des Mat\'eriaux Mol\'{e}culaires, STI-IMX-LOMM, Station 3, Ecole Polytechnique F\'{e}d\'{e}rale de Lausanne, CH-1015, Lausanne, Switzerland.}}

\begin{abstract}
We develop a new three-dimensional multiparticle Monte Carlo ({\it 3DmpMC})
approach in order to study the hopping charge transport in disordered organic
molecular media. The approach is applied here to study the charge transport
across an energetically disordered organic molecular heterojunction, known to
strongly influence the characteristics of the multilayer devices based on thin
organic films. The role of energetic disorder and its spatial correlations,
known to govern the transport in the bulk, are examined here for the bilayer
homopolar system where the heterojunction represents the bottleneck for the
transport. We study the effects of disorder on both sides of the heterojunction,
the effects of the spatial correlation within each material and among the
layers. Most importantly, the {\it 3DmpMC} approach permits us to treat
correctly the effects of the Coulomb interaction among carriers in the region
where the charge accumulation in the device is particularly important and the
Coulomb interaction most pronounced. The Coulomb interaction enhances the
current by increasing the electric field at the heterojunction as well as by
affecting the thermalization of the carriers in front of the barrier. Our MC
simulations are supplemented by the master equation (ME) calculations in order
to build a rather comprehensive picture of the hopping transport over the
homopolar heterojunction.
\end{abstract}

\maketitle

\section{Introduction}

The physical processes governing the devices based on thin organic amorphous
films have been the subject of significant interest during the last
decade.\cite{SomeReview} This includes the charge transport, exciton creation,
and electron-hole recombination in the strong electric field of the order of
$1$MV/cm. These and other processes form the basics of functioning of present and
forthcoming organic electronic devices. In addition, these devices are usually
composed of more than one organic material. This introduces the organic
heterojunctions which often play the major role in the device characteristics.
This shows both experimentally and in some elaborate device model
simulations.\cite{Tang87,TangSlyke,expHJ2,expHJ3,Staudigel,TutisJAP01,HTL03} The
structure of the heterojunction is known to affect the efficiency, electric
characteristics of the device, and its durability.  Therefore the understanding
of the electronic processes at organic heterojunctions is essential for the
proper understanding of the whole device. In spite of that the theoretical work
aimed at understanding the organic heterojunctions, in terms of their
structure, is scarce. This may be contrasted with numerous theoretical
investigations on the effect of disorder and correlations on injection and charge
mobility within the bulk.\cite{GC96}\cite{WAB99} This former research has
established a general picture of electronic transport in disordered molecular
materials and polymers. It proceeds via electron hopping among representative
molecular states. In particular, one deals with the highest occupied molecular
orbital (HOMO) and the lowest unoccupied molecular orbital (LUMO). The occupancy of
HOMO and LUMO orbitals changes as the device is switched on. First, extra
electrons and holes are being introduced from electrodes (injection) onto
neighboring molecules. As the carriers spread further into the bulk they reach
the heterojunctions. The most important characteristic of a heterojunction is
the difference of LUMO and HOMO levels between different molecular species. This difference often 
shows as the energy barrier for the carriers (electrons entering the material
with higher LUMO level or holes entering the material with lower HOMO). This
almost regularly occurs in multilayer structures of organic light emitting diodes (OLED), where lower LUMO
materials near the cathode (facilitating electron injection) meet the higher LUMO
materials inside the device (more appropriate for light emission of the photon
with several eV's in energy). Similarly, one encounters the energy barrier for
holes at heterojunctions closer to an anode. This stepwise injection, with 
intermediate materials between the electrode and the electro-luminescent
material, is known to facilitate the charge inflow into the device. However, it
also happens that in that way the bottleneck for charge transport moves from
the electrodes towards the organic heterojunctions. The latter then dominate the
electric characteristics of organic multilayer structures. It should be pointed
out from the outset that the average value of the energetic barrier at the
heterojunction is not its sole property to determine its permitivity to
carrier flow. It is the energetic and structural disorder that also affect the
performance of the heterojunction. Disorder is expected to modify the
heterojunction characteristics, similarly as it predominantly determines the
carrier mobility in an organic amorphous material. Another direct consequence of
the barrier is the charge accumulation at the heterojunction, which is much more
pronounced there than in the bulk. Therefore the {\it Coulomb effects} are also
more pronounced at the heterojunctions than elsewhere. In that respect it is
important to emphasize that in organic molecular materials the interacting
electrons are well {\it localized} in space, with the localization length being
of the size of a single molecule. Thus the nature of the Coulomb interaction
is very much different from that in band semiconductors, where the spread of the
electron wavefunction is considerably bigger than the mean carrier separation.
The picture of interaction in disordered molecular media resembles the one
between slowly moving "point charges". Therefore the interaction energy depends
essentially on the charge configuration at a given instant of time and may not
be properly expressed through the average occupancy of sites - the effect of
correlations is pronounced.

For the reasons given above the most direct theoretical approach to an organic
heterojunction is to develop a 3D numerical model for the hopping transport of
many particles. In that respect it may be noted that numerical models for 
single-particle hopping were successfully used in the past to examine the effect of
disorder on mobility and injection.\cite{GC96,Bassler93,ConW97,GC95} These
models have been conveniently implemented through the Monte Carlo (MC)
algorithm. The MC approach has been used to simulate the hopping movement of a
particle exploring the energy landscape of the molecular material. In order to
consider the heterojunction problem many such particles have to be considered
concurrently.

It should be noted immediately that the rather simple multiparticle MC approach
proposed here is possible because in the systems under consideration the quantum
phase of the electronic wave function is essentially lost in the 
(phonon-assisted) hop. Therefore, the orbital quantum effects are not expected to show
in the assembly of particles. Instead, one may consider electrons as classical
particles exhibiting hopping transport. On the other hand, the spin state may
be, and certainly {\it is}, in the absence of spin-orbit interaction centers,
remembered in the hop.

There is one additional reason why the multiparticle MC approach is rather
appropriate for studying organic molecular devices. The approach in which each
particle is followed through the system profits from the fact that even for a
large number of molecules (several million in our simulations) the number of
particles is relatively low (several hundred). Indeed, in real systems like
organic molecular diodes the density of particles is usually very low because of
the lack of countercharges. Therefore the Coulomb forces limit the charge
density. On the other hand, when both electrons and holes exist within a single
organic layer, it is the recombination that limits the density. In both cases
the molecules greatly outnumber the carriers. Only in the regions near
bipolar heterojunctions, where both electrons and holes accumulate, the density
of particles may be significantly increased. However, even then the MC approach
may turn tractable.

As it will become apparent, the 3D multiparticle Monte Carlo ({\it 3DmpMC})
approach may be applied to a wide set of problems aside from heterojunctions, e.g.
to study the recombination crossection, for considering the density and doping
effects for the hopping transport, or for considering the influence of the
discrete charge hopping on the space charge effects. As for heterojunctions, it
may be applied both to a heterojunction where only one type of carriers is
present (homopolar) as well as to those where electrons and holes meet and
recombine (bipolar). Multiplying the effects of disorder, correlations,
density effects, Coulomb interactions, recombination and interface structures
brings us to great many number of possibilities, each within the reach of the
method. However, we do not consider all these issues in the present paper. The
paper introduces the {\it 3DmpMC} method and discusses its application to the
homopolar heterojunction problem. We consider the heterojunctions where only
energetic disorder is present, postponing the discussion of the structural
disorder for a separate paper. With all these limitations imposed it still turns
out that the problem of a flat, energetically disordered, homopolar
heterojunction presents a multitude of effects. We believe that all these have
to be understood before addressing those related to the interface structured in
the real space and before considering the bipolar problem.

Confining ourselves to the problem with a single type of carriers we bring in an additional simplification - we do not have to take care about spin degree of freedom. In the simulation we do not
permit a single molecule to be occupied by more than one carrier of a given type
(no two extra electrons or two extra holes may reside on the same molecule
simultaneously), whatever its spin projection. This constraint, imposed in
accordance with the usual treatment of {\it localized} (donor, acceptor) states in
doped semiconductors,\cite{Sapoval} is maintained within the model even when we
switch off the long-range tail (nearest neighbour and beyond) of the Coulomb
interaction. Effectively, the constraint implies that single {\it state} is
available per molecule for a given carrier. This implies that the spin degree of
freedom will not show in the calculation for "electrons-only" or "holes-only"
devices. It is understood that the "spinless-particle" approach may not be
appropriate when electrons and holes are considered concurrently, with the
recombination being dependent on the spin state of the particles.

From here the paper goes as follows. In the next three sections we introduce the
Monte Carlo model and the numerical procedure as well as the master equation
(ME) approach. We will discuss first the MC and ME results for the dependence of
the particle current on disorder strength in variously parameterized
heterojunctions. We then discuss in more detail the effect of long-range and
short-range Coulomb interaction on the barrier crossing. The ratio of the
forward and backward hopping rates is discussed in a separate section. This
ratio also turns out to be important in the final section where we consider in
more detail the effect of the precise form of the hopping law on the properties
of the system.

\section{The model}

As the model of a heterojunction we consider the model of two cubic lattices
placed side by side. For MC simulations we use two lattices of the size
$200\times 200\times 50$ sites. Each of them represents a separate organic layer
(denoted as layer I and layer II). The overall thickness of the model sample (100
molecular monolayers) is thus typical of real devices. The electrodes that are
placed on both sides of the device impose the electric field and the carrier
injection. For the two lateral directions, perpendicular to the applied electric
field, we impose periodic boundary conditions to account for the extension of
the device in those directions.

Here we describe a somewhat more general model than required for the
calculations in the rest of the paper. In the present section we consider the
system where both electrons and holes are present. Each site is then
characterized by two energy levels, LUMO and HOMO. The energetic disorder is
introduced by adding the random values $\epsilon_i$ to the site LUMO and HOMO
energies. These values follow the Gaussian distribution
$(2\pi\sigma^2)^{-1/2}\exp\left(-\epsilon^2/\sigma^2\right)$. We consider both
uncorrelated and correlated disorder. For the latter, the short-range
correlations between the site energies are further imposed by taking the energy
of the $i$-th site to be proportional to the average of the energies of all the
sites contained within a sphere of radius $a_{corr}$, i.e. $\epsilon_i=N_{corr}
\sum_{k}\epsilon_k$, where $N_{corr}$ is a normalization
factor.\cite{GC95,RC00} The correlation in the heterojunction region is
handled separately. We consider both a) the case where the correlations inside
each device layer are independent, and b) the case where the correlations extend
over the organic/organic heterojunction. In the latter case we speak of
"over-barrier" correlations. This case can occur especially when the energetic
disorder is caused by random orientation of molecular dipoles. The correlations,
occurring due to the long range of the dipolar potential, then extend beyond the
interface between different materials.

The charge carriers (electrons and holes) interact through the Coulomb
interaction, $q_iq_j/4\pi \epsilon_0 \epsilon_r r_{ij}$, screened by the
dielectric constant of the material. Of course, the ''self-interaction'' of the
electron is excluded. In particular this means that the energy levels of the two
sites involved in hopping have no Coulomb contributions from the electron
involved. On the other hand, these energy levels do experience the Coulomb
shifts from all other particles in the device. In practice the calculation of
the Coulomb shift is simplified by considering only the particles in the basic
cell as point charges. For a given site this basic cell accounts for the portion
of the device, with lateral extension $200\times 200$ in our case, centered
around the particular site. The Monte Carlo calculations for one-component
plasma\cite{Brush66} showed that this approximation (termed ''the minimum-image
convention''), produces small differences with respect to the exact Ewald sum for
a diluted plasma.

Due to weak binding forces in organic molecular semiconductors the carriers
are localized on molecules. The hopping of electrons and holes from one
localized site to another occurs when (as a result of the electron-lattice
interaction) the atomic vibrations change the relative energy of these localized
states. Three parameters determine the conditions of the charge transfer between
the sites: the transfer integral $J$ related to the wave function overlap
between sites, the coupling strength $g$ of the carrier to the lattice and the
typical phonon energy $\hbar\omega_0$.\cite{Emin92-93,Emin75,ZBP94} In the
organic materials used as transport and emission layers in OLEDs, the transfer
integral $J$ is usually less than $0.1$eV.\cite{FRB04} The phonon mode which
produces the relaxation is the C-C stretching mode which interchanges single and
double bonds of the conjugated backbone ($\hbar\omega_0=0.15$ to $0.2$eV). The
coupling constant can be deduced from the relaxation energy of the molecule upon
charging: $g=E_r/\hbar\omega_0$. Values of the relaxation energies have been
calculated in Alq$_3$\cite{CBA98} and in PPV.\cite{ZBM03} They range between
$0.06$ and $0.3$ eV. A comprehensive review on the relative importance of the
relevant parameters and the qualitative effects that we follow here, has
been published by Emin.\cite{Emin86}

It may be recalled that the ratio $J/\hbar\omega_0$ is informative about the
adiabaticity of the process. In the molecular materials considered here, where
$J$ is small, hopping is usually nonadiabatic: the intermolecular coupling
energy is so small that the electronic carrier can only occasionally respond to
favorable changes of atomic configurations by moving between sites. On  the other
hand, the coupling strength $g$ in those materials is in the range from $0.5$ to
$2$, implying a rather strong coupling. The hopping rate for such a process is
given by\cite{Emin75}
\begin{equation}
\nu_{ij}=\frac{J^2}{\hbar}\sqrt{\frac{\pi}{2E_b
kT}}\exp\left(-\frac{E_b+\epsilon_i-\epsilon_j}{2kT}\right)
\exp\left[-\frac{\left(\epsilon_i-\epsilon_j\right)^2}{8E_b kT}\right],  
\label{polaron:hop}
\end{equation}
where $E_b$ is the polaron binding energy.
   
However, for historical reasons, hopping in molecular materials is usually
considered through formula appropriate for a nonadiabatic process in the weak-coupling regime. The
Miller-Abrahams formula, originally proposed for shallow impurity transport in
semiconductors, gives the jump rate for such a phonon-assisted
process,\cite{MA60}

\begin{equation}
\nu_{ij} = \nu_0 \exp\left(-2\gamma a
\frac{r_{ij}}{a}\right)\exp\left[-\left(\epsilon_i-\epsilon_j+
\left|\epsilon_i-\epsilon_j\right|\right)/2kT\right],
\label{MA:hop}
\end{equation} 
where $\nu_0$ is the attempt-to-jump frequency. Here only the rate upward in
energy is activated.

Powered by the success of the Gaussian Disorder Model of B\"assler and
coworkers\cite{Bassler93} and due to its simplicity, the Miller-Abrahams formula
was used in almost all simulations related to transport in
OLEDs\cite{WAB99,AEB01}. Nevertheless, in spite of their differences, the two
formulae lead to some similar physical processes in a disordered system. Field-dependent mobility is also qualitatively similar
for the two laws. On the other hand, one big difference is the appearance in
Eq.(\ref{polaron:hop}) of the Marcus inverted region. In fact at higher fields,
or more generally when $\left|\epsilon_i-\epsilon_j\right|\gg E_b$, the hopping
rate starts decreasing due to the decreasing probability of phonon emission for a
downward hop in energy.\cite{Emin75}

In order to provide the link with most of the papers on the hopping transport in
the systems with Gaussian disorder, in this work we  use mainly the hopping
processes described by Eq.(\ref{MA:hop}). The question of the influence of
different hopping laws on our conclusions is addressed in the separate
section towards the end of the paper.

\section{The 3D multi particle Monte Carlo algorithm}

The multiparticle Monte Carlo algorithm we use may be roughly described as
follows. Basically, we follow the evolution of many particles, electrons and
holes. As the system evolves through hops, at each step one has to select the
particle that is to hop next. For that purpose we introduce the "dwelling time"
$ t_\mathrm{dwell}^i $. It is an amount of time that the particle at site $i$ spends there
before the hop, $ t_\mathrm{dwell}^i =\left(\sum_{j}\nu_{ij}\right)^{-1}$. The particle with
the smallest dwelling time is selected to perform the jump at a given instant of
time. Once the particle at site $k$ is chosen to perform the next hop, the
destination of the hop is determined by a random number generator in accordance
with the relative magnitudes of the $\nu_{kj}$ of sites inside a box of size
$Nb\times Nb\times Nb$ (e.g. with $Nb=6$). After the hop is performed the time
variable $t$ is advanced by the corresponding dwelling time - i.e. the time step
in the simulation is defined by the dwelling time of the particle executing the
hop.

More precisely, the algorithm goes as follows:
\begin{enumerate}
\item Let $t$ denote the absolute time of the simulation and $t_i$ the 
private time of particle at site $i$. The latter is set to zero when the particle arrives 
at the site and advances when the particle do not hop. 
Initially $t_i$=0 for all sites $i$ populated by particles. 
The total number of occupied sites corresponds to $N$, the number 
of particles (electrons and holes) in the system.

\item Calculate the energy level shifts for the sites inside the hopping box ($Nb\times Nb\times Nb$) 
around each particle. Each energy level is determined by the external electric 
field and the Coulomb potential of the particles in the system.

\item Calculate the hopping probability for each site in the hopping box.

\item Deduce the dwelling time $t_{\mathrm{dwell}}^i$ for all occupied sites.

\item The site $k$, occupied by the particle to hop next, is determined 
by searching for the smallest $\left(t_\mathrm{dwell}^i-t_i\right)$ 
among occupied sites.

\item Choose at random the site $j$ within the hopping box of site $k$ as the 
destination for the hop, obeying the relative probabilities calculated above. . The particle is de-assigned from site $k$ and assigned to site $j$.

\item Update the times: 
$\left\{
\begin{array}{l}
t = t + \left(t_\mathrm{dwell}^k-t_k\right) \\
t_j = 0 \\
t_i = t_i + \left(t_\mathrm{dwell}^k-t_k\right), {\rm for occupied sites},
i\ne j \\
\end{array}
\right. $

\item If $\left(t_\mathrm{dwell}^k-t_k\right) < 0$ then the particle at 
site $k$ hops to the site $j$ chosen in the same manner, while the time variables are updated as follows:

$\left\{
\begin{array}{l}
t_j = 0 \\
t, t_i \;\mathrm{unchanged, for\,  occupied\, sites}\, i\ne j)
\end{array}
\right. $
\end{enumerate}

The new configuration of particles is used to continue the simulation 
by re-entering the above algorithm from step $2$.

The unit of time in the code is given by 
\begin{equation}
t_0 = \left[6\nu_0 \exp\left(-2\gamma a\right)\right]^{-1}
\end{equation}
The parameters $\nu_0$, and $\gamma$ are the same throughout the system. In the simulations presented here $\nu_0=10^{13}\textrm{ s}^{-1}$ and $\gamma=8.33\textrm{ nm}^{-1}$.

As the model is currently set for the study of organic/organic interfaces,  the
details of the exact injection mechanism will not matter very much. However, it  is
important to establish the mechanism for the carrier inflow towards  the
interface. For example, one can use an appropriate injection formula with  the
parameters suitably chosen, e.g. the one discussed in Ref.\cite{ScottCPL1999}. 
The probability for the injection at the sites near the electrode is then determined by the electric field at the 
injecting electrodes. The latter is separately calculated within the code.
Alternatively, one may fix the number of carriers (separately electrons and holes)
in the device and create a carrier at a random position near the injecting
electrode whenever a particle disappears either by leaving the device at the
opposite electrode or in the recombination process. This approach of a fixed
number of particles is chosen for the present paper, because in that case the
total current in the system is the property of the heterojunction and not
related to electrodes. However, to some extent, the results obtained within the
two "injection approaches" may be transformed from one into another. For 
heterojunction investigations this basically reduces to calculating the current
per particle, instead of considering only the value of the current at a fixed 
number of particles. The calculation of the Coulomb effects on the injection
field is simplified by the fact that most of the carriers are located in the
heterojunction area.

The simulations related to the present paper begin with a random distribution of 
carriers. Snapshots of all the variables of interest are taken each 50  hops of 
the fastest carrier (i.e. in periods in which at least one of the carriers makes 
50 hops). The steady state (the same average current along the whole device) is reached within 
the period of approximately 14 snapshots. After this stage  the average of all 
the variables is taken in the time window of 50 snapshots. In this way we obtain 
satisfactory statistics for typical  values of disorder and field strengths.

The algorithm described above has been implemented on a cluster of
Linux-based  machines. The numerical application of the {\it 3DmpMC}  method is
much facilitated  by the development of the computer cluster calculation
techniques, where a handful of particles is taken care of by each computer node. The
information required to calculate the Coulomb interaction between particles is
communicated among nodes using Message Passing Interface (MPI) calls.

\section{The master equation approach}

Many MC methods are known for their conceptual simplicity and a straightforward 
numerical approach. However, they are also known for the need to examine great
many states of a system in order to reduce the statistical error. The present
MC method shares this property. The other method that is sometimes used to
examine the system of particles hopping in a disordered energy manifold is the
master equation (ME) approach. The master equation approach has some advantages
over the MC approach with respect to reducing the statistical uncertainties, as
discussed in Ref.\cite{YuPRBPRL}

On the other hand, the ME approach fails to treat the Coulomb
interaction correctly, both long-range and-short range, although some attempts have been
done to consider
approximately the no-double-occupancy constraint within this approach.\cite{YuPRBPRL,PasveBlomPRL2005} Here we use this approach where
appropriate to supplement our results of the MC simulations. A more detailed
account on the master equation approach may be found in
Refs.\cite{YuPRBPRL,TutisPRB2004} where it was applied for studying
the mobility and injection in a 3D disordered system, respectively. Here we briefly outline
the method and contrast it to the {\it 3DmpMC} approach.

The basic assumption of ME approach is that the occupancy $n_i$ of site $i$
evolves in time according to the equation
\begin{equation}
\frac{dn_{i}}{dt}=\sum_{j.j\neq i}(-n_{i}\nu_{ij}+n_{j}\nu_{ji})
\end{equation}
where $\nu_{ij}$'s are hopping rates already introduced within the MC approach.
For an ensemble of non-interacting particles this is a perfectly rightful
assumption if $n_i$ represents the average occupancy of the site over a time
scale much bigger than the inverse of the characteristic hopping frequency.
Finding the steady state of the system ($dn_i/dt=0$) then amounts to solving the
(huge) linear system of equations for $n_i$'s, each of the equations being of the type as given
above. The source part for the linear systems is provided by suitable additional
terms that describe the coupling to the particle reservoir (electrodes). The
solution simultaneously takes care of all possible hopping processes
in the system. This is at odds with the MC approach which samples only
a fraction of the bonds in the system, and those that are sampled are visited a
finite number of times. However, as usual with the Monte Carlo, the statistics
improves as the sampling extends.

As already mentioned, the effects of interaction are difficult to introduce
correctly within the ME approach. For example, although some neighboring sites
$i$ and $j$ may be occupied to the similar extent on average, $n_i\sim n_j$, in
reality this may come form the same particle visiting both sites in turn. This
means that at each moment one site is occupied while the other is empty and vice versa. Thus,
calculating the interaction energy as the product $n_i\times n_j$ is obviously
erroneous.

The simplest interaction that one wants to take account of is the strong
repulsion among  particles on meeting on the single molecule. If this
interaction is strong enough the double occupancy of a molecule will be
practically forbidden. Thus the particle at a site $i$ will not be able to jump to
a site $j$ if the latter is occupied. In the quantum Hamiltonian language this is
often described by effectively multiplying the hopping operator by
$(1-\hat{n}_j)$, $\hat{n}_j$ being the particle density operator for site $j$.
A similar extension to the master equation approach is obviously approximate
since  $n_j$ in the master equation does not denote an operator but an average site
occupancy.

However, on the qualitative level the extension of the ME equations to
\begin{equation}
\frac{dn_{i}}{dt}=\sum_{j, j \neq i }\left[-n_{i}\nu_{ij}(1-n_{j})+n_{j}\nu_{ji}(1-n_{i})\right]
\end{equation}
seems advantageous. In the first place it assures that, irrespective of the
number of carriers in the system (which should be less than the number of molecular sites,
of course), the system will acquire a steady state with $n_i<1$. For example,
this prevents deep states (traps) to be overpopulated when the average number of
carriers increases. Moreover, in the limiting case of no sources or sinks in the
system, the equilibrium will be given by
$n_i=\{1+\exp[(E_{i}-\mu)/k_{B}T]\}^{-1}$, with $\mu$ set by the total number of
particles. The latter result is what one expects for a site with
no-double-occupancy constraint being in equilibrium with the reservoir
represented by the rest of the system. For these reasons the preceding
generalization of the ME seems suitable for considering a system with a finite
concentration of particles. On the technical level, the extension is paid off by
the fact that the steady-state equations become nonlinear. This requires
devising a suitable mathematical procedure to solve the huge nonlinear system of
equations for a given number of particles in the system. On the fundamental
level, it must be kept in mind that the nonlinear ME approach is intrinsically
approximate in the sense of the mean-field approximation, as explained above.
Therefore, the multi-particle Monte Carlo approach, while possibly inferior
(statistics-wise) for considering a system of non-interacting particles, seems to be an appropriate and superior approach for considering a system of real, interacting particles.

All applications of the ME approach in this paper are thus for particles that do
not interact with long-range Coulomb forces. The linear and nonlinear ME
solutions are compared to assess the importance of short-range repulsion for
a given number of particles in the system. To reduce the complexity of the
computational problem we consider only the nearest-neighbor hopping in all ME
calculations. The results are generally consistent with MC simulations, given
the strong decay of $\nu_{ij}$'s with increasing distance among molecular
sites. All ME calculations presented here are related to systems with correlated
disorder in both layers, the correlation not extending over the heterojunction.

\section{The effect of disorder on the current through the heterojunction}

\begin{figure}[!h]
\begin{center}
\includegraphics[]{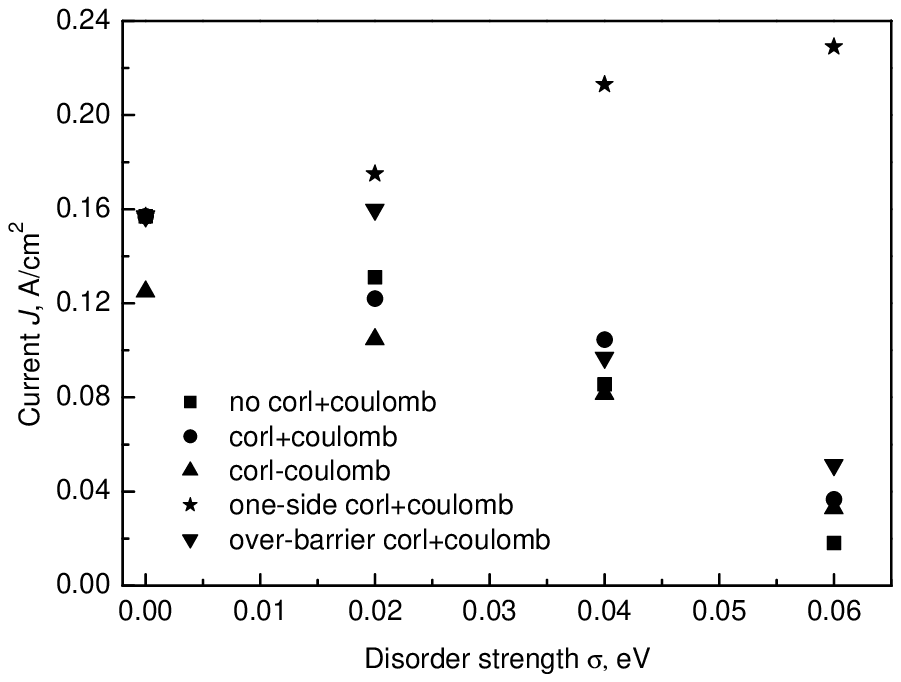}
\caption{}
\label{nwfig01.eps}
\end{center}
\end{figure}

The first important result of our MC simulations for homopolar heterojunctions
are collected in Fig. \ref{nwfig01.eps}. The figure shows the results for the
current in the systems with correlated and uncorrelated disorder, with and
without long-range Coulomb interactions, and contrasts the role of disorder in
one vs. both layers in the bilayer device. For layers with correlated disorder
the correlation length is fixed to three lattice constants.

We briefly comment on the main features discernible from
Fig.\ref{nwfig01.eps}, deferring a more detailed discussion of each of them for
separate sections. The figure shows the general dependence of the barrier
permittivity on disorder strength. The only case where the current increases
with increasing disorder strength is the one in which disorder is confined to
the layer II while the layer I is kept ordered. The observed increase of
the current with disorder is presumably due to the particular points where the
barrier across the heterojunction is locally lowered. The current over the
barrier is generally expected to depend exponentially (i.e. strongly nonlinearly)
on the barrier height. Hence it is expected that the contribution of those
crossing points to the average overcomes the effect of the points where the
barrier is locally increased.

For all other cases the disorder strength is kept equal in both layers. In all
these cases the current diminishes as the strength of disorder increases. Thus
the effect of disorder in the layer I is to {\it reduce} the current. This
effect always dominates the previously observed effect of disorder in the layer II.
As discussed later in the paper and already observed in Ref. \cite{AEB01}, the
carriers accumulated in front of the barrier tend to thermalize to low energy
states there, thus effectively increasing the barrier at the
heterojunction. The figure also suggests that correlation within separate layers is
not particularly beneficial for barrier crossing. On the other hand, the barrier
crossing is facilitated when the correlation among layers is present. It may be
further observed that switching on the long-range Coulomb interaction always
increases the current in the system. This is chiefly because the Coulomb
forces effectively lower the barrier. However, as will be shown soon, the size
of this effect is much smaller than expected from the usual naive argument.

\section{The effect of the Coulomb interaction on the barrier crossing}
\subsection{Long-range Coulomb forces}

\begin{figure}[!h]
\begin{center}
\includegraphics[]{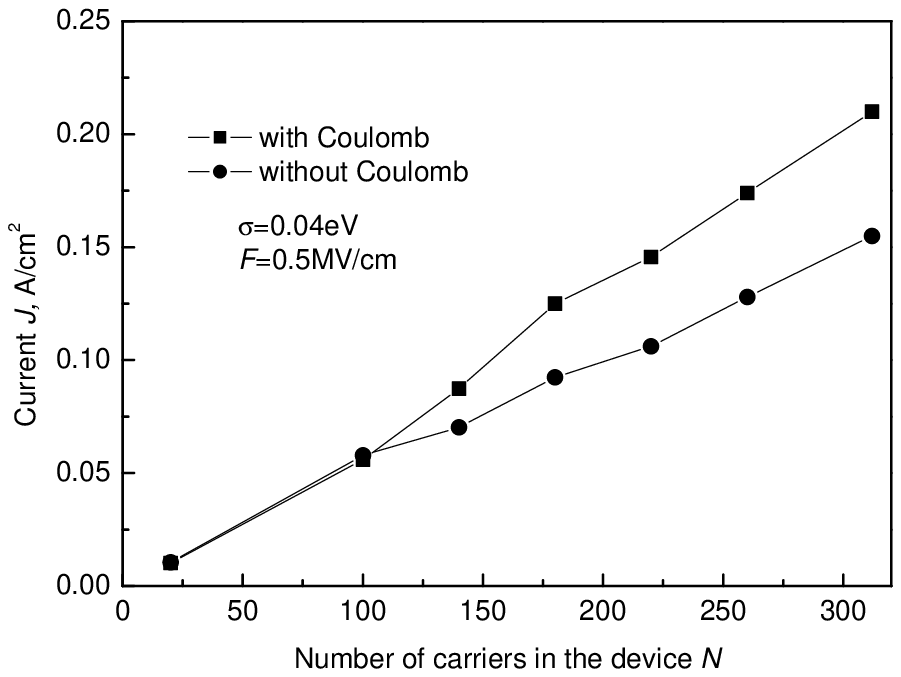}
\caption{}
\label{nwfig04.eps}
\end{center}
\end{figure}

The effect of the interaction may be observed in our simulation as we increase
the number of particles in the system. This is illustrated in
Fig.\ref{nwfig04.eps}. The figure shows the current as the function of the
number of particles in the system for particular values of the barrier height
and disorder strength. The points related to the system of particles without
long-range Coulomb interaction follows a straight line as the number of
particles increases. The effect of the long-range Coulomb interaction is
unobservable for small number of particles in the system. As expected, the
effect starts to show as the number of particles increases.

It is also expected that long-range Coulomb interaction among electrons
accumulated at the heterojunction will help the carriers to pass the barrier.
This is expected since the electric field due to the interaction may lower the
barrier in a way similar to how the external electric field $F$ lowers the barrier by
the amount $Fqa$. The usual straightforward estimate for the barrier lowering
due to particle interaction goes as follows. Let us denote the average planar
charge density at the heterojunction by $\rho_S=qN_s/a^2$, where $N_s$ stands
for the average number of carriers per cell cross section $a^2$. If the charge
at the heterojunction is assumed to be smeared homogeneously over the
heterojunction plane, the extra electric field produced by this charge would be
$F_{\mathrm{hom}}=\rho_S/2(\varepsilon_0 \varepsilon_r)$. Twice that much is the
change of the electric field over the heterojunction. The naive estimate for
barrier lowering is $F_{\mathrm{hom}}qa$.

\begin{figure}[!h]
\begin{center}
\includegraphics[]{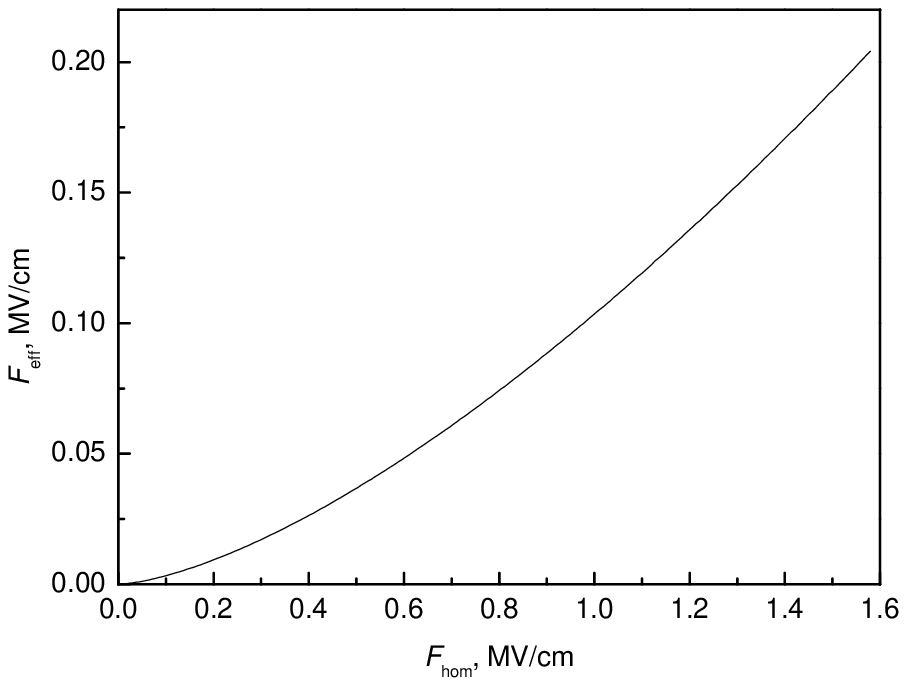}
\caption{}
\label{nwfig06.eps}
\end{center}
\end{figure}

However, it must be recalled that the carrier wavefunction is not at all
homogeneously smeared in the heterojunction plane. The carriers localized on 
molecular sites are spaced by $d \sim \sqrt{1/N_s}$ on average. It is the
Coulomb repulsion among accumulated electrons that tends to order them in a
regular mesh in the heterojunction plane. The continuum approximation
$F_{\mathrm{hom}}$ for the electric field is valid only for distances $\delta x
\gg d$ from the heterojunction. At shorter distances the effect of charge
discreteness must be accounted for. Specifically, the distance to be considered
for barrier lowering is $\delta x \sim a$. As the characteristic case for a
homopolar heterojunction is $a \ll d$, the total change of the electric field at
the heterojunction is expected to be smaller than the average field in the
smeared charge picture. The effective electric field responsible for barrier
lowering may be estimated e.g. by assuming that repelling charges order
approximately in the regular square lattice. The effective electric field then
turns out to be very much smaller than $F_{\mathrm{hom}}$, as showed in Fig.
\ref{nwfig06.eps}. 

\begin{figure}[!h]
\begin{center}
\includegraphics[]{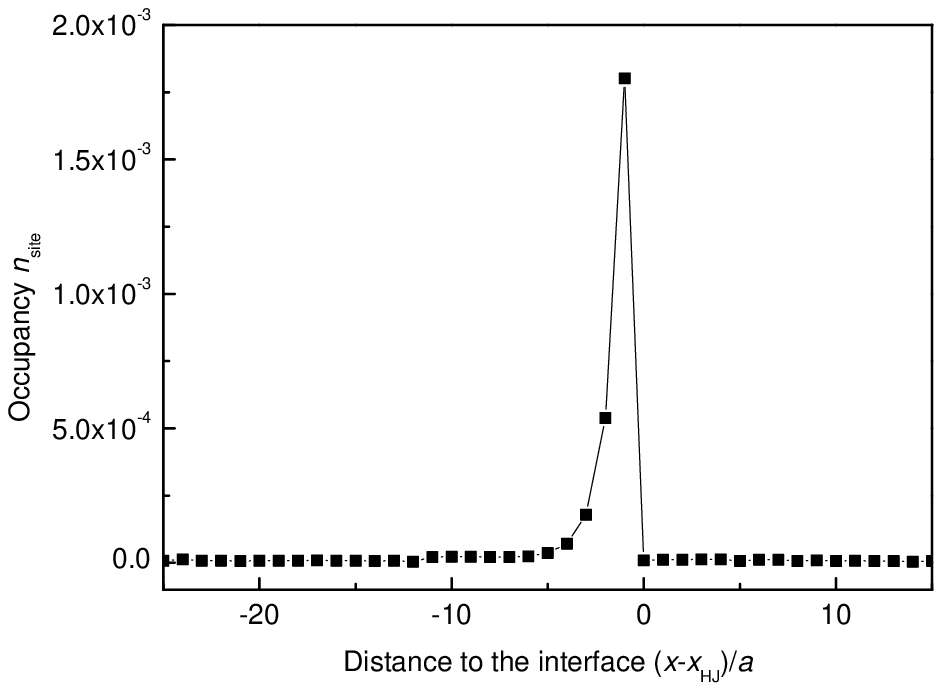}
\caption{}
\label{nwfig07.eps}
\end{center}
\end{figure}

We illustrate these considerations with the case of 140 particles in our $100
\times 200 \times 200$ system. The particles accumulate in front of the
heterojunction, as shown in Fig. \ref{nwfig07.eps}. The planar particle density
per unit cell cross section is $140/(200\times 200)\approx 0.002$, with most of
the particles concentrated in the last few monolayers of the layer I. The surface charge
density is $\rho_S=140q/(200a)^2$. In the continuum approximation the
accumulated charge produces the extra field of approximately 0.25 MV/cm (taking
$\varepsilon_R=3.5$). This field is already quite strong. Twice as many particles
would double the electric field in layer II with respect to the imposed $0.5$ MV/cm
and reduce the electric field in the layer I to zero. For that reason the number of
particles in the system may hardly go much above this concentration.
Further particles would cause a negligibly small or even negative electric field
at the injecting electrode. Taking the calculated extra field of $0.25$ MV/cm as
responsible for barrier lowering would imply a barrier shift of $0.015$ eV. This
would imply a current amplification of approximately ($\exp(0.015{\rm
eV}/k_{B}T)\approx 1.8$) 80 percent. However this is not what one observes in
the MC simulation where the amplification of a few percent is observed.

On the other hand, taking into account the discrete nature of the charge
distribution in the heterojunction plane gives a much smaller electric field. It
turns to be approximately 20 times smaller than the one obtained for the
homogeneously charged plane. The barrier reduction due to particle interaction
is much smaller than obtained above, as well as the current amplification which is then estimated to
$(\exp\left[(0.015 {\rm eV}/20)/k_BT\right]\approx 1.03$) a few percent, as
observed in the simulation.

In summary, the Coulomb effect usually calculated for a homopolar heterojunction
is strongly overestimated. This is true for most multilayer device models, as
well as for some recent analyses of the dominant device
physics.\cite{BlomRecentHJ} On the other hand, the effect of discreteness that
reduces the Coulomb effects in homopolar heterojunctions is expected to
strongly amplify the Coulomb effects among electrons and holes at bipolar
junctions.

\subsection{The hard-ball repulsion}

The short-range repulsion is always present in our {\it 3DmpMC} model.
Therefore its effect can not be deduced directly from the results on the
current density. On the other hand, we did the calculation with and without
no-double-occupancy term within the ME approach. We found that the current
enhancement due to short-range repulsion at a given particle density is less than
one percent for the parameters given in Fig.\ref{nwfig01.eps}.

The effects of the short-range repulsion can be assessed even within the MC
approach by considering the distribution in energy of the particles accumulated
in front of the energy barrier. The particles in the sample spend much more time
in front of the barrier then elsewhere. The number of hops per unit time among
sites within that monolayer greatly outnumbers those that go down-field. This
makes it possible for particles to explore the sites in front of the
barrier rather well and establish a local quasi-equilibrium.

The properties of the quasi-equilibrium with and without
the hard-ball repulsion may be approximately calculated in a simplified manner, and the results may be compared against the MC results. The calculation goes as follows. 
For non-interacting ({\it ni}) particles, the probability of a site being occupied 
is given by the Boltzmann expression $f_B(E)=\exp(-E/k_BT+\mu_{ni}/k_BT)$, where
$\mu_{ni}$ stands for the local chemical potential, which depends on local
concentration of particles. The distribution of the sites in energy, around zero
average energy, being given by $P_0(E)=(2\pi\sigma)^{-1/2}\exp(-E^2/2\sigma^2)$,
the probability distribution for the energy $E$ representing an occupied site
goes as
\begin{equation}
O_{ni}(E)= P_0(E) f_B(E) =\frac{1}{\sqrt{2\pi\sigma^2}}
\exp\left[-\frac{(E+\sigma^2/k_BT)^2}{2\sigma^2}+\frac{\mu_{ni}}{k_BT}+
\frac{\sigma^2}{2k_B^2T^2} \right]
\end{equation}
Thus the mean energy of the occupied state is shifted by $-\sigma^2/k_BT$ with
respect to the case without disorder. This well-known shift of the energy of
occupied states due to thermalization\cite{Bassler93} tends to increase the
effective barrier at the heterojunction. 

\begin{figure}[!h]
\begin{center}
\includegraphics[]{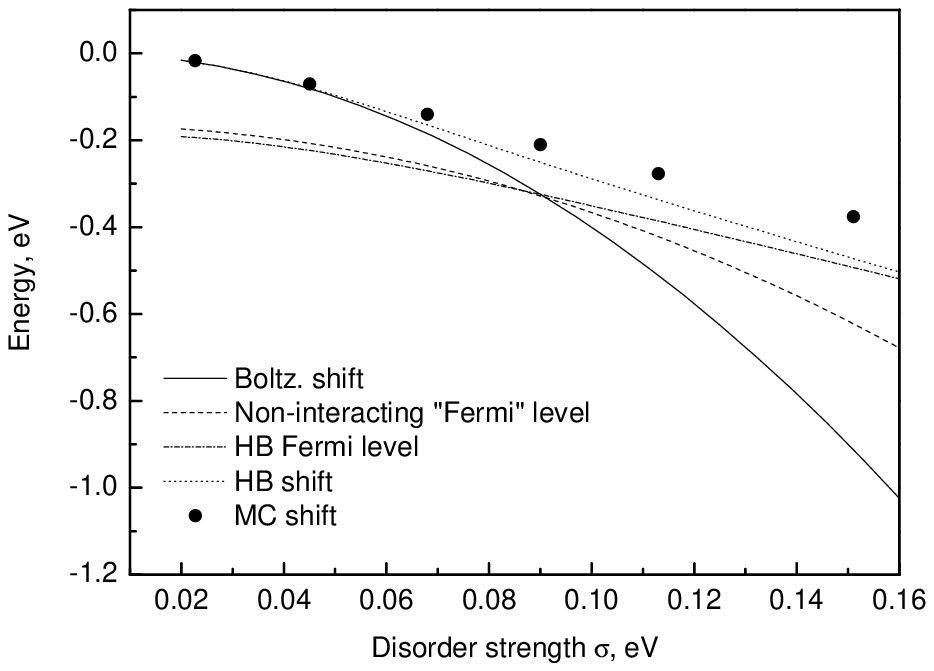}
\caption{}
\label{nwfig09.eps}
\end{center}
\end{figure}

This shift {\it decreases} to some
extent once the hard-ball ({\it hb}) repulsion is taken into account. 
The no-double-occupancy constraint changes the probability of the state of energy
$E$ being occupied to
$f_{hb}(E)=\left\{1+\exp\left[(E-\mu_{hb})/k_BT\right]\right\}^{-1}$,
i.e. to the Fermi-Dirac distribution function.\cite{FDnote}. The chemical
potential and mean energy have to be calculated numerically for this probability distribution. The result for two cases (i.e. with and without hard-ball repulsion) is illustrated in Fig. \ref{nwfig09.eps}. In general, the
hard-ball repulsion {\it reduces} the shift of the mean particle energy due to
disorder. The interaction also shifts the chemical potential upwards.  The MC results are also shown in  Fig. \ref{nwfig09.eps}. The thermalisation obtained within MC simulations is somewhat weaker than the one predicted by simple hard-ball statistics calculation. We believe that this
discrepancy may be due to the confining effect of the electric field. By pushing the carriers towards the interface, the field prevents them from
exploring a larger energy manifold with lower energy levels.
By switching on and off the long-range Coulomb
interaction we checked that those contribute negligibly
to thermalisation at a given particle density.  

Fig. \ref{nwfig09.eps} also shows that for a particular particle density
and for room temperature, the difference between the thermalization shifts is very small for disorder strengths less than 0.06 eV. The difference 
becomes pronounced in systems with stronger disorder or at higher particle
density at the heterojunction. Since the value for $\sigma$ in some disordered organic molecular
material may be rather big (e.g. $\sigma\sim 0.15$ eV for Alq$_3$),
\cite{MalliarasAlq3sigma}, the effect of short-range repulsion is realistic to
expect for some heterojunctions. On the other hand, the particle density at 
homopolar heterojunctions may be much higher only if the operational field is raised much above 0.5 MV/cm. This is because, the operational
field limits the particle surface density at the heterojunction, as already pointed above. 

\section{Backward and forward hopping rates}

Recently Arkhipov et al. got interested in the ratio of backward ($b$) to
forward ($f$) hopping rates at the heterojunction\cite{AEB01}. Their physical
motivation was to resume the hops that lead to injection at a heterojunction
($\propto b-f$) from those that may lead to immediate recombination with a
carrier of the opposite polarity. While their argument may certainly need to be
amended by considering the Coulomb interaction among carriers of opposite
polarity in the latter case, the $b/f$ ratio is indeed important when
considering the permitivity of the barrier to current flow.

\begin{figure}[!h]
\begin{center}
\includegraphics[]{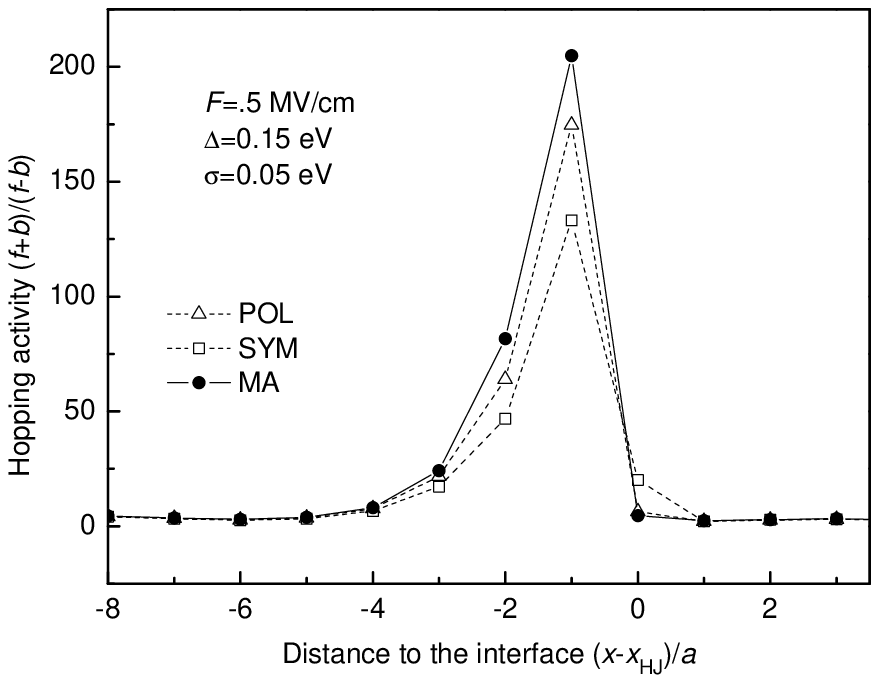}
\caption{}
\label{nwfig10.eps}
\end{center}
\end{figure}

Other motivation to consider the $b/f$ ratio comes from reflecting on the aging
processes in organic devices. A decade ago it was shown by Tang and Van Slyke
that the aging rate of OLED's is proportional to the operating
current.\cite{TangSlyke} This rule, also termed the "Coulomb aging" rule, may
suggest that the hopping rate at the microscopic scale is directly related to
degradation\cite{D-Berner}. However, while the average current density is
constant throughout the device, it is not necessarily so for the hopping rate.
This is because at any given point in the device both forward and backward hops
contribute to the local hopping rate ($\propto f+b$), while the current, being
constant along the device, is proportional to their difference ($\propto f-b$).
The hopping  activity, here defined as $(f+b)/(f-b)=(1+b/f)/(1-b/f)$ is
therefore expected to be dependent on position in an inhomogeneous device, and
especially when considering the region near the heterojunction.
Fig.\ref{nwfig10.eps} shows this dependence as obtained in our ME calculation.
It may be observed that the hopping activity is many times bigger near the
heterojunction than elsewhere. However, it is not the very heterojunction where
the hopping activity is maximal. The hopping activity maximizes in the last few
monolayers of the layer I, just before the barrier. This is mostly due to
the increased concentration of particles in that region and the relatively low probability
for further down-field hops.

\begin{figure}[!h]
\begin{center}
\includegraphics[]{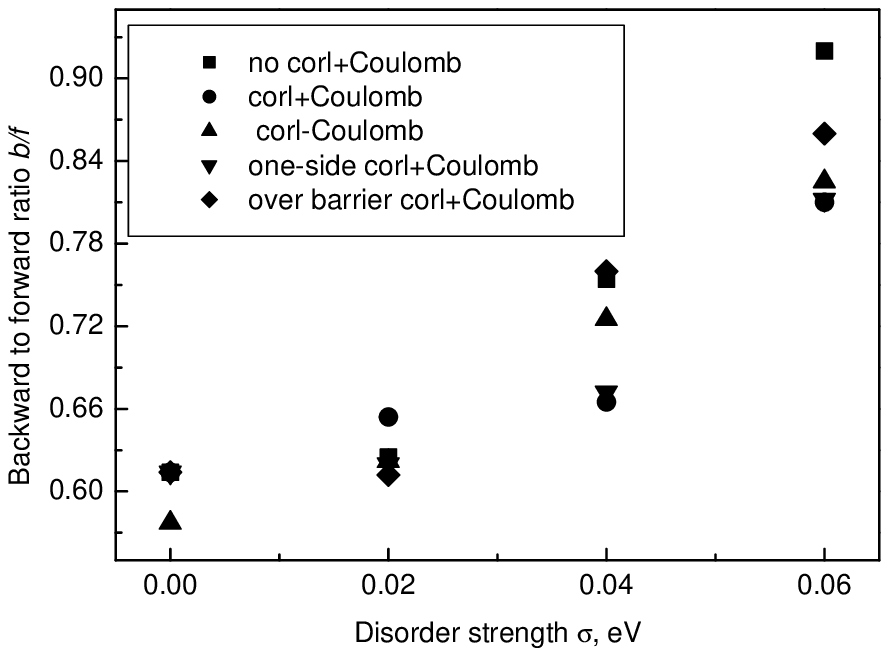}
\caption{}
\label{nwfig11.eps}
\end{center}
\end{figure}

\begin{figure}[!h]   
\begin{center}
\includegraphics[]{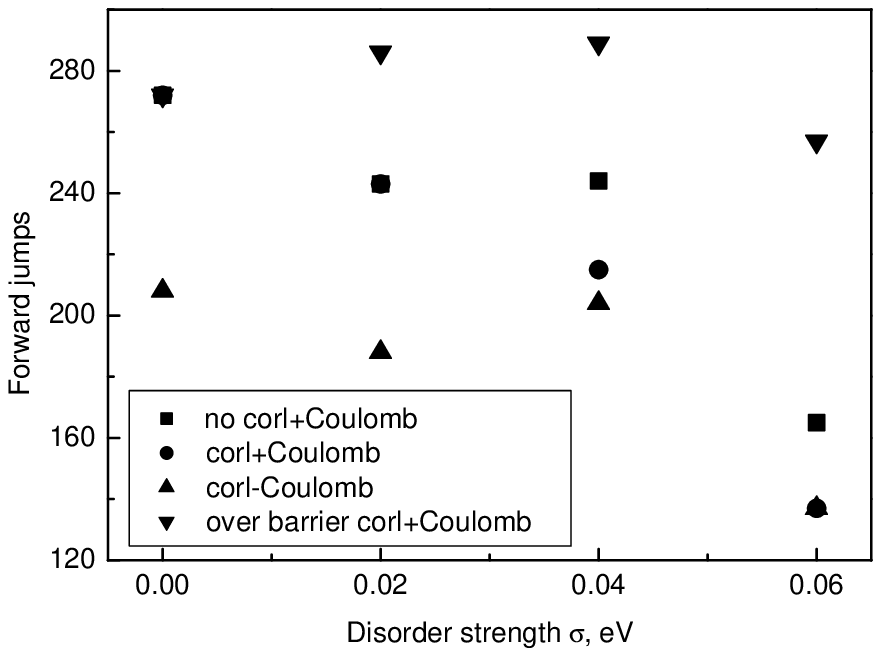}
\caption{}
\label{nwfig12.eps}
\end{center}
\end{figure}

Concentrating on the hops over the heterojunction, which determine the
heterojunction permittivity to particle flow, we investigated the effect of the
structure of the heterojunction and the Coulomb interaction on the $b/f$ ratio.
The result is shown in Fig. \ref{nwfig11.eps}. While generally the $b/f$ ratio
increases with the increasing strength of disorder, one may also notice the
effect of correlations and the Coulomb interaction. First, correlation generally
lowers the $b/f$ ratio, especially as disorder increases. The effect is
possibly related to the fact that the carriers that enter the layer II face
a locally more ordered environment, which  implies smaller probability for
returning hops. Second, the Coulomb interaction increases the relative
importance of backward jumps to some extent, especially at low disorder strength. We do not
have a simple explanation for the latter effect. Considering the
forward hopping rate separately in Fig. \ref{nwfig12.eps}, we find that it increases on
switching on the Coulomb interaction, to the effect of increasing the current, in
spite of an increased $b/f$ ratio, $J\propto f\times(1-b/f)$. However, the increase
of the $b/f$ ratio cannot be explained simply in terms of increased effective
electric field at the interface. This will become clear shortly as we consider
the dependence of $b/f$ ratio on the electric field.

For the nearest-neighbor hopping and without long-range Coulomb interaction, an
analytic expression for the $b/f$ ratio can be derived in the limit of zero
disorder. The derivation starts from the fact that the density in an ordered system is
constant throughout layer II. This establishes the relation between the current
and the particle density in the same layer. The current may be also expressed by
the difference between forward and backward hopping rates at the very
heterojunction. Then, assuming $\Delta>Fqa$, the result for the M.-A. hopping
formula follows,
\begin{equation}
\left(b/f\right)_{\mathrm{HJ}}=1/\left[2-\exp(-Fqa/k_B T)\right]
\label{eq:7}
\end{equation}
The zero-disorder value of $b/f$ is $0.59$ for the parameters of Fig.
\ref{nwfig11.eps}, in accordance with MC results.

\begin{figure}[!h]
\begin{center}
\includegraphics[]{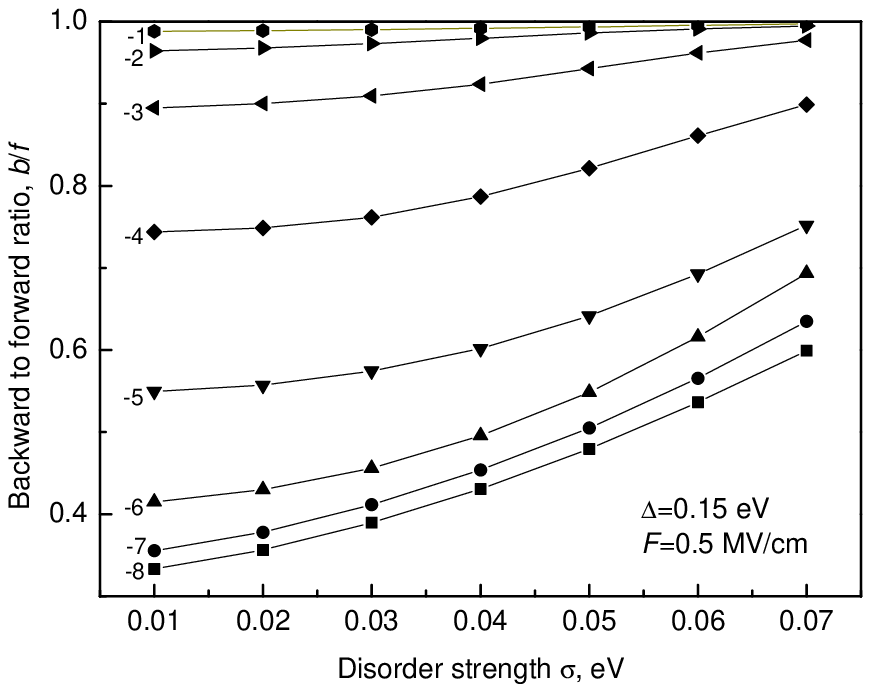}
\caption{}
\label{nwfig13.eps}
\end{center}
\end{figure}

\begin{figure}[!h]
\begin{center}
\includegraphics[]{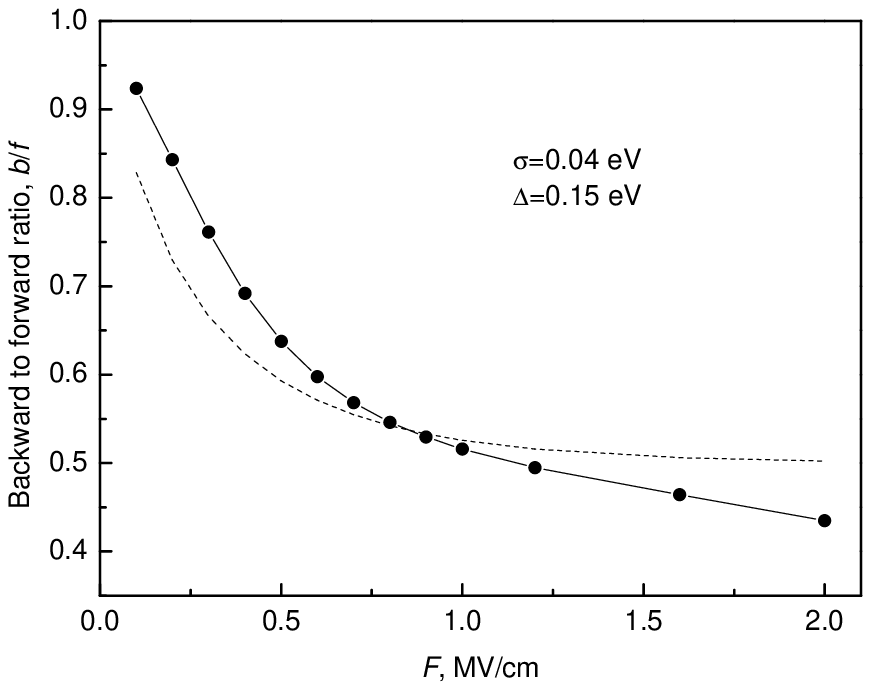}
\caption{}
\label{nwfig15.eps}
\end{center}
\end{figure}

As for the dependence on disorder strength $\sigma$, the MC results are
supplemented here by the ME results for $b/f$ away from the heterojunction
(Fig.\ref{nwfig13.eps}). The dependence on the electric field given in
Fig.\ref{nwfig15.eps}, while roughly following the analytical result for
$\sigma=0$, changes differently with disorder for low and high electric field,
the crossover being roughly defined by $Fqa\sim\Delta/3$.

\section{The effect of the hopping law on the barrier crossing}

\begin{figure}[!h]
\begin{center}
\includegraphics[]{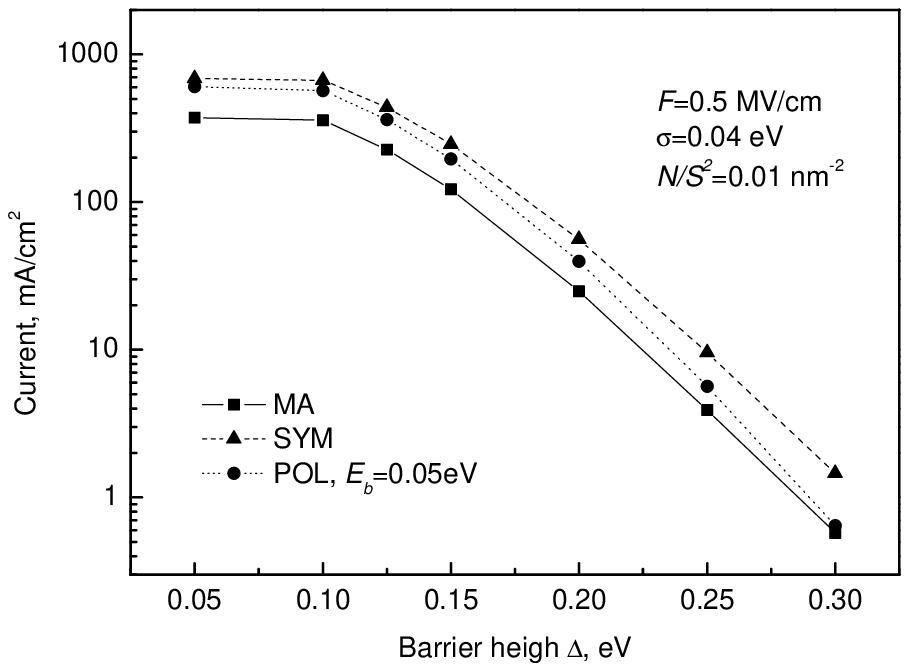}
\caption{}
\label{nwfig16.eps}
\end{center}
\end{figure}

One of the questions that arise when discussing the effects in a system with
hopping conduction is the qualitative dependence of these effects on the precise
form of the hopping law. While the results presented above were obtained by
assuming the Miller-Abrahams hopping law, qualitatively similar results are obtained
if one uses the symmetric hopping law or the small-polaron hopping formula. This
is illustrated in Fig.\ref{nwfig16.eps} (ME calculations). The figure shows
the current through the heterojunction as a function of the barrier height
$\Delta$, at a given disorder strength. The hopping laws used for this
calculation are
\begin{eqnarray}
\nu_{ij}^{MA}&=&\nu_{nn}\exp\left(-\frac{E_{i}-E_{j}+\left|E_{i}-E_{j}\right|}{k_{B}T}\right), \\
\nu_{ij}^{SYM}&=&\nu_{nn}\exp\left(-\frac{E_{i}-E_{j}}{2k_{B}T}\right), \\
\nu_{ij}^{POL}&=&\nu_{nn}\exp\left[-\frac{E_{i}-E_{j}}{2k_{B}T}-\frac{\left(E_{i}-E_{j}\right)^{2}}{8k_{B}TE_{b}}\right].
\end{eqnarray}
To simplify the comparison, the bare value for the nearest-neighbor hopping frequency $\nu_{nn}$ was chosen the same for
all three hopping laws, while the hopping was confined to the nearest neighbors. The
number of carriers in the system was also kept the same.

For $\Delta=0$ the current through the system reflects the value of the mobility
in a homogeneous system for different hopping laws. The issue of mobility
vs. hopping law in a disordered system has been addressed in the
literature.\cite{NovikovPRL98} It was shown numerically that the Miller-Abrahams
and the symmetric formula give qualitatively the same field dependence of mobility,
as far as the disorder strength is bigger than the potential drop among
neighboring molecules due to the electric field, $\sigma > F q a$. While sharing the common
 field dependence, the value for mobility $\mu_{MA}(F)$ in that region is found
generally smaller than $\mu_{SYM}(F)$. We are not aware of any similar considerations for heterojunctions, either numerical or
analytical. Fig. \ref{nwfig16.eps} shows the
dependence of the current through the junction on the barrier height. It shows that the
dependence is similar for all three hopping laws. Although the polaron binding term
reduces the current somewhat as the barrier height increases, the dependences
shown by all three curves are qualitatively similar.

What may seem particularly strange is the fact that the ratio of
$J_{SYM}/J_{MA}$ changes very little over the range where the factor $\exp\left(\Delta/2k_{B}T\right)$, relating different 
hopping laws at the heterojunction, changes
over three and a half orders of magnitude. This may be even stranger when
the distributions of the carriers in space are examined for the three hopping laws. 
No significant differences can be observed - almost all the carriers are grouped
in the first few monolayers in front of the barrier, as illustrated in
Fig.\ref{nwfig07.eps}. Therefore the forward hopping rate at the heterojunction
is approximately $\exp\left[(\Delta-Fqa)/2k_{B}T\right]$ times bigger for
the symmetric hopping law then for the Miller-Abrahams (MA) hopping law. Hence the
backward hopping rate in the symmetric case has to compensate 
the forward hoping rate rather precisely in order to bring the current to the same order of
magnitude as for the MA case.

Fortunately, these considerations may be supported by exact analytic
expressions for a limiting case of the heterojunction between two ordered
layers. As already pointed out in the previous section, in that case the ratio
of backward to forward hopping rates may be derived analytically.\cite{bfMAnote}
For the Miller-Abrahams hopping rate the result was given already in
Eq.\ref{eq:7} whereas for the symmetric law it is
\begin{equation}
\left(\frac{b}{f}\right)_{HJ,SYM}=\frac{1}{1-e^{-\Delta/2k_{B}T}\left(e^{Fqa/k_{B}T}-1\right)},
\end{equation}
Neglecting the minor differences between carrier distributions, the ratio of
currents in the two cases becomes
\begin{equation}
\frac{J_{SYM}}{J_{MA}}\approx
e^{(\Delta-Fqa)/2k_{B}T}\frac{1-(b/f)_{HJ,SYM}}{1-(b/f)_{HJ,MA}}.
\end{equation}
The analytical expressions for $b/f$'s can be substituted into this equation. It  is then readily verified  that the big difference in forward hop frequencies between the two cases is compensated by respective
backward hop rates. This analytic expression for $J_{SYM}/J_{MA}$ also 
reproduces approximately the value obtained numerically (Fig.\ref{nwfig16.eps}) for finite
disorder. We believe that this qualitative result also stands in the core of the
aforementioned similarity between mobilities for the two hopping laws in
homogeneously disordered systems. In the latter case the bottlenecks for
transport are again barriers bigger then $Fqa$, statistically generated and
scattered in space. The argument relating the two hopping laws is still expected to
apply on those barriers.

\section{Conclusion}

In summary, we presented a new multiparticle Monte Carlo method that is suitable
for addressing a number of presently unexplored questions related to electronic
transport in organic disordered materials. In this publication we applied the
method to a homopolar heterojunction, considering the influence of disorder and Coulomb
interaction on the current through the heterojunction. Supplementing our MC
results with the master-equation (ME) calculations, we developed a rather detailed
insight into the effects of short-range and long-range Coulomb repulsion,
forward vs. backward hopping processes and the precise form of the hopping law.
In particular, we showed the estimates based on a continuum approximation strongly overestimate the effect of the long-range Coulomb interactions on the
current through the heterojunction. Also, the long-range Coulomb interactions do not affect visibly
the thermalisation of carriers in front of the barrier. It is rather the
no-double-occupancy constraint which dominantly determines the effective barrier
height. The presence of disorder and especially the energy barrier at the
heterojunction lead to position-dependent forward and backward hopping rates.
The frequency of charging and discharging of molecules then varies throughout
the device. This likely affects the aging and the degradation. As in the case of
bulk transport, the microscopic hopping law appeared to play a minor role in
heterojunction barrier crossing provided that the polaron binding energy is
smaller than the energy barrier. This study presents basis for further
theoretical studies of more complicated (mixed, graded, rough) homopolar
heterojunctions that are being developed experimentally. The multiparticle MC
method introduced in this paper is currently applied for studying the
recombination processes at bipolar heterojunctions. 

\begin{acknowledgments}

We thank Dr. D. Berner for many helpful discussions. We are also grateful to Dr. B.
Horvati{\'c} for discussions and remarks on the manuscript which significantly improved the paper. The MC calculations
in this paper were done on "Janus", an HP SC45 cluster at DIT-EPFL. The ME
calculations were preformed on the "Grozd" Opteron cluster at the Institute of
Physics, partly using the PARDISO library.\cite{refPARDISO}

\end{acknowledgments}

\newpage

\section*{Figure captions}

\begin{list}{}{}

\item{\bf Figure 1}: (MC) Current across the heterojunction as a function of
disorder strength. Following cases are considered: {\it no corl+Coulomb}: non
correlated disorder and (long-range) Coulomb interactions included; {\it
corl+Coulomb}: correlated disorder within each layer and Coulomb interactions
included; {\it corl-Coulomb}: correlated disorder within each layer and without
long-range Coulomb interaction; {\it one-side corl+Coulomb}: correlated disorder
in the second layer while the first layer is ordered, the Coulomb interactions
are included; {\it over-barrier correlated +Coulomb}: the correlations in the
energy disorder present in both layers extending over the heterojunction,
long-range Coulomb forces being included in the calculation. The number of
particles in the system ($ 100 \times 200 \times 200$ sites) is always 140. The
bare barrier height (the difference among average LUMO/HOMO energy of two
materials) at the heterojunction is 0.15 eV. The strength of externally applied
electric field is 0.5 MV/cm.

\item{\bf Figure 2}: (MC) The dependence of the current in the system on the
number of particles in the system ($100 \times 200 \times 200$ sites). The
squares stands for the results where the interaction among particles is full
Coulomb interaction. The lower curve (filled circles) stands the current in the
system where only "hard-ball" short-range interaction is taken into account
(i.e. double occupancy of a site is forbidden but no long-range repulsion is
accounted for). The imposed electric field is $0.5$ MV/cm. The bare barrier
height is $\Delta=0.15$ eV.

\item{\bf Figure 3}: The effective field at the homopolar heterojunctions
leading to barrier lowering, as produced by the accumulated charges, vs. the
electric field of the equivalent homogenously smeared planar charge.
$F_{\mathrm{eff}}$ is estimated here by assuming that electrons at the
heterojunction are ordered as point charges forming a regular square mesh.

\item{\bf Figure 4}: (MC) The average occupancy of the molecular site in the
system as a function of the position of the site with respect to the plane of
the heterojunction (the first monolayer of the layer II is positioned at zero). The
density in front of the barrier is much bigger than in the bulk of both layers.
From the heterojunction to the left the occupancy falls approximately
exponentially. The model parameters are: $\sigma=0.04$eV, $\Delta=0.15$eV,
$a=0.6$nm, $F=0.5$MV/cm. The data represent the results for for 140 particles in the system of
$100\times 200 \times 200$ sites.

\item{\bf Figure 5}: The energy shift due to thermalisation in front of the
barrier, shown as the function of disorder strength. {\it Boltz. shift} is for
non-interacting particles when the Boltzman law gives the orbital occupancy.
{\it HB shift} is the calculation for the particles with hard-ball interaction
in the equilibrium for the average molecular occupancy of $2\times 10^{-3}$. The
filled circles is what we get from the MC simulations.

\item{\bf Figure 6}: (ME) The hopping activity in the system is shown as the
function of the position along the system. Here $f$ and $b$ stand for
the forward and backward hopping frequency between successive monolayers, respectively. $f+b$ is
the average frequency of hops in both direction, while $f-b$ is constant throughout
the system, being proportional to the particle current in the field direction. Three
graphs correspond to three hopping rules (MA: Miller-Abrahams, SYM: symmetric
hopping law, POL: small polaron hopping formula with polaron binding energy
$E_b=0.05$ eV). $\sigma=0.05$ eV, $a=0.6$ nm, $\Delta=0.15$ eV. Backward hops at
the heterojunction are the most pronounced for the symmetric hopping law, while less
pronounced for the Miller-Abrahams and small-polaron hopping formulae.

\item{\bf Figure 7}: (MC) Variation of the ratio of backward to forward hopping
rates with disorder. $F=0.5$ V/cm. $\Delta=0.15$ eV. The number of particles in
the system is 140.

\item{\bf Figure 8}: (MC) Number of forward hops as a function of disorder. The
forward hopping rate is strongly increased by long-range Coulomb forces,
especially at low disorder. It is also increased by the disorder correlations
between layers ({\it "over barrier corl"}).

\item{\bf Figure 9}: (ME) The ratio of the backward to forward hop frequency as
the function of the strength of disorder. Different graphs refer to 
different positions with respect to the barrier, as indicated by the labels (e.g.
-3 denotes the position that is three monolayers before the barrier). The
spatial dependence of $b/f$ is not at all pronounced in layer II. Throughout the
layer II the $b(\sigma)/f(\sigma)$ curve resembles the one labeled by -8, 
corresponding to the monolayer of layer I that is  rather far from the
heterojunction.

\item{\bf Figure 10}: (ME) The ratio of backward and forward hopping frequency at
the very barrier as a function of the electric field at the barrier. The dashed
line represents the result without disorder, described by the formula in the
text.

\item{\bf Figure 11}: (ME) The dependence of the current through the junction as
the function of the barrier height. The three curves represent the results for three
different hopping laws. The density of particles per device cross section,
$140/(200*0.6 {\rm nm})^2 \approx 0.01 nm^{-2}$, is the same in all three cases.

\end{list}
   
\newpage

\begin{thebibliography}{999}
\bibitem{FDnote} The Fermi-Dirac distribution for the hard-ball problem is the consequence of considering spinless particles. Each site then contributes to the density of states with a single state. Considering the particles with spin one-half one would have two (2) states per site (one per spin projection). In that case the probability for the site being occupied would read as $1/\left\{1+\frac{1}{2}\exp\left[(E-\mu_2)/k_BT\right]\right\}$, where 2 comes from the number of spin projections. It may be easily checked that for a given concentration of particles the two approaches (with and without spin) to the hard-ball interaction give the same distribution. The factor $\frac{1}{2}$ in the second approach is compensated by the difference in the chemical potential with respect to the spinless approach.
\end{thebibliography}
\end{document}